\documentclass[prb,twocolumn,superscriptaddress,showpacs]{revtex4}
\usepackage{times}
\usepackage{amsmath}
\usepackage{amsthm}
\usepackage{amssymb}
\usepackage{amsbsy}
\usepackage{graphicx}
\usepackage{pstricks}

\begin{document}

\bibliographystyle{apsrev}

\title{Multiferroicity induced by dislocated spin-density waves}

\author{Joseph J. Betouras}
\affiliation{School of Physics and Astronomy,
Scottish Universities Physics Alliance, 
University of St Andrews, North Haugh KY16 9SS, UK.}
\affiliation{Instituut-Lorentz for Theoretical Physics, Leiden University,
P.O. Box 9506, 2300 RA Leiden, The Netherlands.}
\author{Gianluca Giovannetti}
\affiliation{Instituut-Lorentz for Theoretical Physics, Leiden University,
P.O. Box 9506, 2300 RA Leiden, The Netherlands.}
\affiliation{Faculty of Science and Technology, University of Twente, P.O. Box
217, 7500 AE, Enschede, The Netherlands.}
\author{Jeroen van den Brink}
\affiliation{Instituut-Lorentz for Theoretical Physics, Leiden University,
P.O. Box 9506, 2300 RA Leiden, The Netherlands.}
\affiliation{Institute for Molecules and Materials, Radboud Universiteit
Nijmegen, P.O. Box 9010, 6500 GL Nijmegen, The Netherlands.}

\date{\today}

\begin{abstract}

We uncover a new pathway towards multiferroicity, showing how magnetism can drive ferroelectricity without relying on inversion symmetry breaking of the magnetic ordering. Our free-energy analysis demonstrates that any commensurate spin-density-wave ordering with a phase dislocation, even if it is collinear, gives rise to an electric polarization. Due to the dislocation the electronic and magnetic and inversion centers do not coincide, which turns out to be a sufficient condition for multiferroic coupling. The novel mechanism explains the formation of multiferroic phases at the magnetic commensurability transitions, such as the ones observed in YMn$_{2}$O$_{5}$ and related compounds.  We predict that in these multiferroics an oscillating electrical polarization is concomitant with the uniform polarization. On the basis of our theory we put forward new types of magnetic materials that are potentially ferroelectric. 

\end{abstract}

  \pacs{
    77.80.-e,
    75.30.Fv,   
    75.47.Lx  }
\maketitle

{\it Introduction} The recent discovery of novel materials in which magnetic and ferroelectric order can coexist --termed {\textit{multiferroics}}-- has sparked a new surge of interest in this field~\cite{eerenstein,cheong,kimura,hur,lottermoser,vanaken,efremov,efremov2,hemberger,cohen,ederer}. From a technological point of view the possibility to control magnetic properties by electric fields and, vice versa, ferroelectric order by magnetic fields, is very desirable. By definition multiferroics, sometimes also called magnetoelectrics, possess at least two switchable states of electric polarization ($\bf {P})$ and magnetization ($\bf {M})$. Although there is a number of materials that exhibit both ferroelectricity and magnetism, it might come as a surprise that there need
not necessarily be a large coupling between them. Typically the ferroelectric transition temperature is much higher than the magnetic one and a coupling between the two order parameters is weak. Representative examples of this behavior are the transition metal perovskites BiFeO$_3$ and BiMnO$_3$  \cite{khomskii,hill,ederer2,dossantos}. 

For potential applications of multiferroics it is very important to establish general mechanisms that give rise to a coupling of ferroelectricity and magnetism. More specifically, one would like to construct situations where the two ordering temperatures can become close or even coincide: in that case one expects strong multiferroic behavior with a pronounced interdependence of the ferroelectric and magnetic order parameters. This then directly opens up the possibility to switch the macroscopic ferroelectric polarization by applying an external magnetic field, which is the key to future applications of multiferroic materials.

In this Letter we report a novel route for potential strong coupling multiferroicity. A straight forward Ginzburg-Landau free-energy analysis demonstrates that spin-density-wave (SDW) ordering with a phase dislocation gives rise to an electric polarization. A remarkable feature is that in such a SDW a commensurate oscillating electric polarization causes a net uniform ferroelectric moment and that incommensurate ordering does not. Therefore our mechanism explains the formation of multiferroic phases at the magnetic commensurability transitions, such as the one observed in YMn$_{2}$O$_{5}$ --indeed a dislocated SDW material with an acentric magnetization. Based on this novel mechanism we will also suggest new types of multiferroic materials.

From a theoretical point of view the question of possible magneto-electric couplings is best addressed within the framework of the Landau theory of phase transitions since it is based on symmetry considerations alone.  In this sense it captures the essence of any microscopic mechanism for the coupling between ferroelectricity and magnetism \cite{chandra}. Such a Landau analysis has already uncovered a case in which ferroelectricity is induced by magnetic ordering. This is the case when the magnetic ordering is of a type that breaks chiral symmetry. This is realized in spiral or helical spin-density wave systems \cite{dzyaloshinskii,mostovoy,lawes,sergienko,katsura}. For instance TbMnO$_{3}$, DyMnO$_{3}$, Ni$_{3}$V$_{2}$O$_{8}$ and the recently discovered multiferroic MnWO$_4$ all fall in this class \cite{kimura,khomskii,lawes,goto,kimura2,taniguchi}. ``Ferroelectricity generated by magnetic chirality'' has established itself as the leading paradigm for both theoretical and experimental investigations in the field of multiferroics with magnetically induced ferroelectricity.

However, a ``rule'' that chiral symmetry needs to be broken in order to induce a ferroelectric moment at a magnetic phase transition is 
questionable. First of all, there are notable exceptions, in particular the manganites RMn$_{2}$O$_{5}$ (R is a rare earth, e.g. Tb, Ho, Dy or Y). 
Experiment shows that in these materials a ferroelectric polarization emerges that is due to collinear magnetic ordering --with no indication of magnetic chiral symmetry breaking \cite{blake,chapon}. 
Second, in several of these systems the ferroelectric polarization appears spontaneously at magnetic commensurate-incommensurate transitions. 
This has even lead to the claim that --on empirical grounds-- a commensurate spin state be essential to the ferroelectricity in multiferroic RMn$_{2}$O$_{5}$ \cite{kimura3}. 
Such an empirical rule cannot be explained within the theoretical framework of chiral symmetry breaking. Instead, as we show below this behavior is inherent to a novel type of multiferroics 
which displays a strong interdependence of magnetization and ferroelectric polarization: materials with a dislocated, acentric spin-density wave ordering.
 
{\it Free-energy analysis} We prove the above assertions by considering a continuum field theory of the Landau  type that incorporates at the same time ferroelectric polarization $\bf {P}$ and magnetic spin-density wave (SDW) order $\bf {M}\left( {\bf {r}} \right)$, where $\bf {r}$ is the spatial coordinate. 
The form of the coupling between the electric polarization $\bf {P}\left( {\bf {r}} \right)$ and magnetization $\bf {M}\left( {\bf {r}} \right)$ can be found from general symmetry arguments \cite{dzyaloshinskii,mostovoy}. First, time reversal, $t \to  - t$, should leave the magnetoelectric coupling 
invariant. As it transforms $\bf M \to  - \bf M$ and leaves $\bf P$ invariant, time reversal requires that the lowest order coupling has to be quadratic in $\bf M$. The symmetry of spatial inversion, i.e., $\bf r \to  - \bf r$, which sends $\bf P$ to 
$- \bf P$ and leaves $\bf M$ invariant, is respected when the coupling of a homogeneous polarization to an inhomogeneous magnetization is linear in $\bf P$ and contains one gradient of $\bf M$. These symmetry considerations lead to a 
magneto-electric coupling term in the Landau free energy of the form

\begin{eqnarray}
F_{ME} (\bf r) &=& \bf P \cdot \left( \gamma  \bf \nabla ({\bf M}^2 )  \right. \nonumber \\
&&+ \gamma '  \left[ {\bf M}(\bf \nabla  \cdot \bf M)   \right.   - \left. \left.  (\bf M \cdot \bf  \nabla )\bf M \right] + ... \right).
\end{eqnarray}

The first term on the right hand side is proportional to the total derivative of the square of the magnetization. It only gives a surface contribution when the free energy is integrated over the spatial coordinates and the polarization $ \bf P$ is assumed to be independent of $\bf r$ \cite{dzyaloshinskii, mostovoy}. It is central to our physical ideas \textit{not} to make this assumption. 

It is easy to show that the second term in the equation above, proportional to $\gamma '$, is only nonzero if the magnetization $\bf M$ breaks chiral symmetry, which is the canonical route towards a strong dependence of $\bf P$ on $\bf M$. On physical grounds this term can readily be understood. In a ferroelectric, inversion symmetry is broken as the system sustains a macroscopic polarization $\bf P$ that is pointing into a particular direction in space. Therefore this polarization can only couple to the magnetization if and only if ${\bf M}$ also has a directionality and lacks a center of inversion symmetry. One immediately understands that this occurs when the magnetization is spiraling along some axis. This also implies, vice versa, that the chiral magnetic ordering induces a ferroelectric polarization. Recently this situation was considered in detail \cite{mostovoy}, we do not consider it here, set $\gamma ' = 0$ in the following and only consider the case of non-zero $\gamma$.

As our focus is on systems that in absence of magnetism show no instability to ferroelectricity, we only take into account the quadratic term in $\bf P$ 
in the electric part of the free energy:
\begin{equation}
F_E ({\bf r}) =  \frac{\bf P\left( {\bf r} \right)^2 } {2\chi _E \left( {\bf 
r} \right)}
\end{equation}
where $\chi _E \left( {\bf r} \right)$ is the dielectric susceptibility. From the variation of $F_{ME} ({\bf r}) + F_E (\bf r)$ with respect to $\bf P\left({\bf r} \right)$ for a given magnetization one can find the value of the static polarization field. It is important to note that the polarization is due to all charges in the system: the atomic nuclei in the lattice plus the electrons. Magnetization, instead, is a density wave of the spin of the electrons alone. The dependence of the fields $\bf P\left( {\bf r} \right)$ and $\bf M\left( {\bf r} \right)$ on their spatial coordinates ${\bf r} = \left( {x,y,z} \right)$ need not be the same, of course. As we will consider an acentric SDW ordering, we explicitly keep the spatial dependence of the dielectric susceptibility in the equation above. For simplicity we shall keep only the $x$-dependence, as our results are readily generalized to include the other spatial dependencies. 

{\it Dislocated SDW}
The magnetization of an phase dislocated, acentric SDW is given by $M = M_0 \cos (q_m x + \phi )$, where 
$q_m $ is the magnetic ordering wavevector and $\phi $ its phase. The fact that the phase of the magnetization is dislocated implies that $\phi $ is finite: the magnetization is shifted with respect to the electronic density. Hence the magnetization is also phase-dislocated with respect to the lattice. However, as the magnetization is collinear and sinusoidal, it obviously has a center of symmetry and no directionality. Thus on the basis of the arguments that exist prior to the present work, one would be led to conclude that it does not couple to $\bf P$. Such a conclusion, however, would be incorrect. In our theory for multiferoicity we indeed assume both the electronic/lattice structure and the magnetic ordering to be inversion invariant, but due to the dislocation {\it the electronic and magnetic inversion centers do not coincide}. The implies that magnetization and polarization {\it together} break inversion symmetry, which turns out to be a sufficient condition for multiferroic coupling. This underlines the interdependence of the lattice and magnetic structure for generating a ferroelectric moment.  

The expression of the free energy can be simplified by first introducing in Eq. (2) the Fourier components of the inverse electronic
susceptibility (we only need and keep the first two: $\chi _E^{ - 1}  = e_0  + e_1 \cos (qx)$, which is assumed to be positive definite) and by using the ansatz 
$P = p_0  + p_1 \cos (qx)$. Note that the oscillating part of the polarization, proportional to  $p_1 $, does not contribute to a net ferroelectric moment as its spatial
average vanishes. However, its presence is crucial, as this oscillating polarization couples directly to the gradient of the magnetization squared and,
at the same time to the uniform polarization $p_0 $.

{\it Electric polarizations} We integrate the free energy over its spatial coordinates and  minimize it with respect to $p_0 $ and $p_1 $.  
This directly leads to our main result for the homogenous ferroelectric polarization:
\begin{equation}
p_0  = \frac{ - \gamma q_m M_0^2 }{2} \frac{e_1 }{2e_0^2  - e_1^2 } \sin 2\phi . 
\end{equation}
We also find a concomitant oscillating polarization in the ferroelectric equilibrium state 
\begin{equation}
p_1  =  - {2e_0 p_0 } / {e_1 }, 
\end{equation}
which is particular to our magnetoelectric coupling mechanism.
These expressions hold under the condition that $q_m  = q/2$, i.e. the magnetic wavevector is \textit{commensurate} and equal to half of the lattice 
wavevector. The polarization vanishes for the case that $q_m $ is incommensurate. These relations between the two ordering wavevectors $q$ and
$q_m$ are due to the fact that the spatial integral of the free energy terms of the kind $\int cos(qx) cos(q'x) \ dx$ vanishes unless $q=q'$. From the expression above 
it is clear that a finite dislocation of the SDW with respect to the electronic density --a non-zero value of $\phi $ -- is essential to produce a non-zero macroscopic ferroelectric moment \cite{betouras}. 

The physical explanation for the resulting ferroelectricity is elementary. As mentioned above, our acentric magnetization density ${\bf M}^2$ is sinusoidal and by itself inversion invariant (we can always choose an appropriate origin of the coordinate system). Therefore it does {\it not} possess an {\it absolute} directionality. However, in an acentric SDW system ${\bf M}^2$ is lagging behind somewhat with respect to the polarization $\bf P$, which is the immediate consequence of the finite phase difference $\phi$. Thus we conclude that ${\bf M}^2$ has a directionality {\it relative} to $\bf P$, which is a sufficient condition for a direct coupling between the two order parameters and the emergence of a macroscopic ferroelectric polarization in the acentric SDW systems.  

{\it Application to materials} The above analysis can be directly applied to the above mentioned class of multiferroic compounds $R$Mn$_{2}$O$_{5}$. These insulators order antiferromagnetically below 45 K and display a complex sequence of incommensurate (ICM) and commensurate (CM) magnetic phases and no sign of magnetic chirality. However, it has been known since a long time that for instance YMn$_{2}$O$_{5}$ displays acentric SDW ordering and significant magnetoelastic effects \cite{chapon,wilkinson}. The acentricity of the SDW is due to the frustration of magnetic order \cite{chapon}. From our calculations one expects a ferroelectric polarization to appear if the acentric SDW ordering is commensurate, with a polarization that is along the direction of $\nabla({\bf M})^2$. Indeed experimentally in YMn$_{2}$O$_{5}$ a spontaneous ferroelectric polarization along the crystallographic $b$-axis appears {\it when commensurate spin ordering sets in} below 45 K. Most remarkably, at 23 K a magnetic CM to ICM transition takes place. At this transition the magnitude of the magnetization does not change, but the ferroelectric polarization collapses, in perfect agreement with our theoretical considerations. 

The theoretical observation that ferroelectricity depends directly on the commensurability of the dislocated SDW is also supported by thermal expansion data for $R$=Ho, Dy and Tb. These show that also the high temperature ferroelectric transition around 45 K coincides with a lock-in of the magnetic wave-vector from incommensurate to commensurate values \cite{delacruz}. In HoMn$_{2}$O$_{5}$, moreover, at the low temperature ferroelectric transition a magnetic field induces ICM-CM transitions that go hand in hand with the appearance of macroscopic
polarization \cite{kimura3}. All these experimental data are in accordance with the acentric SDW route to multiferroicity that we propose in this Letter. Note that in ICM phases one can expect some residual ferroelectricity to appear when the system is partially commensurable or if domain formation occurs.

Another class of systems to which our analysis applies are the perovskite manganites e.g. Pr$_{1-x}$Ca$_{x}$MnO$_{3}$. For large doping concentration a number of different charge, spin and orbital ordered states are predicted and found~\cite{brink1,brink2}. In particular, there is experimental evidence that for a phase of the system that exhibits a spin ordering of the manganese moments that is centered on Mn-Mn \textit{bonds}~\cite{daoud}. Theoretically an acentric SDW is found to be the 
magnetic groundstate when 0.4$<$x$<$0.5 and indeed instability of the lattice towards a polar ferroelectric distortion is expected and found \cite{efremov,krueger}. In single crystals of another member of the same class, Gd$_{1-x}$Sr$_{x}$MnO$_{3}$  electric polarization has been recently observed as well \cite{kadomtseva} and it is to be attributed to the same physics. 

So the present novel magnetoelectric coupling explains the observed multiferroic behavior in two classes of ceramic materials. Our theory suggests that a third, completely different class of systems can show this behavior: organic charge transfer salts. Quasi one-dimensional molecular crystals (TMTTF)$_{2}$Br, (TMTSF)$_{2}$PF$_{6}$ and $\alpha $-(BEDT-TTF)$_{2}M$Hg(SCN)$_{4}$ show an unusual coexistence of spin and charge-density waves \cite{mazumdar}. As in the above mentioned perovskite manganites, such a superposition of spin and charge density wave orderings produces an acentric magnetization. The fact that our conditions of acentricity and commensuration are fulfilled, makes the organic charge transfer salts from the perspective of our theory, candidates {\it par excellence} to exhibit multiferroic behavior. The strength of the induced polarization will depend on the microscopic details of these systems. For a reliable estimate one will need to go beyond the phenomenological Ginzburg-Landau approach that we have presented here~\cite{giovannetti}.

{\it Conclusions} We have shown how magnetism can drive ferroelectricity without relying on inversion symmetry breaking of the magnetic ordering. Starting from the Landau-Ginzburg theory, we have revealed a new mechanism that leads to magnetoelectric coupling in SDW systems. The two key ingredients are the finite dislocation and the commensuration of the SDW with respect to the lattice. There are at least two classes of materials that exhibit this behavior and are thus understood in the light of the present theory. Besides explaining the observed magnetoelectric coupling in these materials we also predict that in these systems an oscillating electrical polarization is concomitant with the uniform polarization and that the magnetic ordering wavevector is half the electronic one. Furthermore, on the basis of our theory we suggest that magnetoelectric coupling occurs in a class of materials that was little considered so far, the organic charge transfer salts, opening up a general pathway to construct new multiferroic materials. 

{\it Acknowledgments}
We thank Maxim Mostovoy, Tsuyoshi Kimura, Claude Pasquier, Laurent Chapon, Paolo Radaelli and Jan Zaanen for stimulating discussions. This work is supported by FOM and the Dutch Science Foundation.

\end{document}